\documentclass[twocolumn,times]{aastex61}

\def\be{\begin{equation}}
\def\ee{\end{equation}}

\newcommand{\cqg}{Class. Quantum Grav.}
\newcommand{\lrr}{Living Rev. Relativ.}

\newcommand{\PlanckHnaughtViewingNinetyULTidalNoSpins}{28}
\newcommand{\RiessHnaughtViewingNinetyULTidalNoSpins}{36}
\newcommand{\add}[1]{\textcolor{black}{#1}}

\submitjournal{ApJ}

\shorttitle{Search for High-energy Neutrinos from GW170817}
\shortauthors{Albert et al.}

\begin{document}

\title{Search for High-energy Neutrinos from Binary Neutron Star Merger GW170817 with ANTARES, IceCube, and the Pierre Auger Observatory}

\AuthorCollaborationLimit=3000
\author{A.~Albert}
\affil{GRPHE - Universit\'e de Haute Alsace - Institut universitaire de technologie de Colmar, 34 rue du Grillenbreit BP 50568 - 68008 Colmar, France}

\author{M.~Andr\'e}
\affil{Technical University of Catalonia, Laboratory of Applied Bioacoustics, Rambla Exposici\'o, 08800 Vilanova i la Geltr\'u, Barcelona, Spain}

\author{M.~Anghinolfi}
\affil{INFN - Sezione di Genova, Via Dodecaneso 33, 16146 Genova, Italy}

%\author{G.~Anton} %also in IceCube
%\affil{Friedrich-Alexander-Universit\"at Erlangen-N\"urnberg, Erlangen Centre for Astroparticle Physics, Erwin-Rommel-Str. 1, 91058 Erlangen, Germany}

\author{M.~Ardid}
\affil{Institut d'Investigaci\'o per a la Gesti\'o Integrada de les Zones Costaneres (IGIC) - Universitat Polit\`ecnica de Val\`encia. C/  Paranimf 1, 46730 Gandia, Spain}

\author{J.-J.~Aubert}
\affil{Aix Marseille Univ, CNRS/IN2P3, CPPM, Marseille, France}

\author{J.~Aublin}
\affil{APC, Univ Paris Diderot, CNRS/IN2P3, CEA/Irfu, Obs de Paris, Sorbonne Paris Cit\'e, France}

\author{T.~Avgitas}
\affil{APC, Univ Paris Diderot, CNRS/IN2P3, CEA/Irfu, Obs de Paris, Sorbonne Paris Cit\'e, France}

\author{B.~Baret}
\affil{APC, Univ Paris Diderot, CNRS/IN2P3, CEA/Irfu, Obs de Paris, Sorbonne Paris Cit\'e, France}

\author{J.~Barrios-Mart\'{\i}}
\affil{IFIC - Instituto de F\'isica Corpuscular (CSIC - Universitat de Val\`encia) c/ Catedr\'atico Jos\'e Beltr\'an, 2 E-46980 Paterna, Valencia, Spain}

\author{S.~Basa}
\affil{LAM - Laboratoire d'Astrophysique de Marseille, P\^ole de l'\'Etoile Site de Ch\^ateau-Gombert, rue Fr\'ed\'eric Joliot-Curie 38,  13388 Marseille Cedex 13, France}

% ...

\collaboration{({\sc{ANTARES}} Collaboration)}

\author{M.~G.~Aartsen}
\affiliation{Department of Physics, University of Adelaide, Adelaide, 5005, Australia}
\author{M.~Ackermann}
\affiliation{DESY, D-15738 Zeuthen, Germany}
\author{J.~Adams}
\affiliation{Dept.~of Physics and Astronomy, University of Canterbury, Private Bag 4800, Christchurch, New Zealand}
\author{J.~A.~Aguilar}
\affiliation{Universit\'e Libre de Bruxelles, Science Faculty CP230, B-1050 Brussels, Belgium}
\author{M.~Ahlers}
\affiliation{Niels Bohr Institute, University of Copenhagen, DK-2100 Copenhagen, Denmark}
\author{M.~Ahrens}
\affiliation{Oskar Klein Centre and Dept.~of Physics, Stockholm University, SE-10691 Stockholm, Sweden}
\author{I.~Al~Samarai}
\affiliation{D\'epartement de physique nucl\'eaire et corpusculaire, Universit\'e de Gen\`eve, CH-1211 Gen\`eve, Switzerland}
\author{D.~Altmann}
\affiliation{Erlangen Centre for Astroparticle Physics, Friedrich-Alexander-Universit\"at Erlangen-N\"urnberg, D-91058 Erlangen, Germany}
\author{K.~Andeen}
\affiliation{Department of Physics, Marquette University, Milwaukee, WI, 53201, USA}
\author{T.~Anderson}
\affiliation{Dept.~of Physics, Pennsylvania State University, University Park, PA 16802, USA}
% ...

\collaboration{(IceCube Collaboration)}

% created on 2017-09-13

\author{A.~Aab}
\affiliation{IMAPP, Radboud University Nijmegen, Nijmegen, The Netherlands}

\author{P.~Abreu}
\affiliation{Laborat\'orio de Instrumenta\c{c}\~ao e F\'\i{}sica Experimental de Part\'\i{}culas -- LIP and Instituto Superior T\'ecnico -- IST, Universidade de Lisboa -- UL, Lisboa, Portugal}

\author{M.~Aglietta}
\affiliation{Osservatorio Astrofisico di Torino (INAF), Torino, Italy}
\affiliation{INFN, Sezione di Torino, Torino, Italy}

\author{I.F.M.~Albuquerque}
\affiliation{Universidade de S\~ao Paulo, Instituto de F\'\i{}sica, S\~ao Paulo, SP, Brazil}

\author{J.M.~Albury}
\affiliation{University of Adelaide, Adelaide, S.A., Australia}

\author{I.~Allekotte}
\affiliation{Centro At\'omico Bariloche and Instituto Balseiro (CNEA-UNCuyo-CONICET), San Carlos de Bariloche, Argentina}

\author{A.~Almela}
\affiliation{Instituto de Tecnolog\'\i{}as en Detecci\'on y Astropart\'\i{}culas (CNEA, CONICET, UNSAM), Buenos Aires, Argentina}
\affiliation{Universidad Tecnol\'ogica Nacional -- Facultad Regional Buenos Aires, Buenos Aires, Argentina}

\author{J.~Alvarez Castillo}
\affiliation{Universidad Nacional Aut\'onoma de M\'exico, M\'exico, D.F., M\'exico}

\author{J.~Alvarez-Mu\~niz}
\affiliation{Universidad de Santiago de Compostela, Santiago de Compostela, Spain}

\author{G.A.~Anastasi}
\affiliation{Gran Sasso Science Institute (INFN), L'Aquila, Italy}
\affiliation{INFN Laboratori Nazionali del Gran Sasso, Assergi (L'Aquila), Italy}

% ...

\collaboration{(The Pierre Auger Collaboration)}

\author{B.~P.~Abbott}
\affiliation{LIGO, California Institute of Technology, Pasadena, CA 91125, USA}
\author{R.~Abbott}
\affiliation{LIGO, California Institute of Technology, Pasadena, CA 91125, USA}
\author{T.~D.~Abbott}
\affiliation{Louisiana State University, Baton Rouge, LA 70803, USA}
\author{F.~Acernese}
\affiliation{Universit\`a di Salerno, Fisciano, I-84084 Salerno, Italy}
\affiliation{INFN, Sezione di Napoli, Complesso Universitario di Monte S.Angelo, I-80126 Napoli, Italy}
\author{K.~Ackley}
\affiliation{University of Florida, Gainesville, FL 32611, USA}
\affiliation{OzGrav, School of Physics \& Astronomy, Monash University, Clayton 3800, Victoria, Australia}
\author{C.~Adams}
\affiliation{LIGO Livingston Observatory, Livingston, LA 70754, USA}
\author{T.~Adams}
\affiliation{Laboratoire d'Annecy-le-Vieux de Physique des Particules (LAPP), Universit\'e Savoie Mont Blanc, CNRS/IN2P3, F-74941 Annecy, France}
\author{P.~Addesso}
\affiliation{University of Sannio at Benevento, I-82100 Benevento, Italy and INFN, Sezione di Napoli, I-80100 Napoli, Italy}
\author{R.~X.~Adhikari}
\affiliation{LIGO, California Institute of Technology, Pasadena, CA 91125, USA}
\author{V.~B.~Adya}
\affiliation{Max Planck Institute for Gravitational Physics (Albert Einstein Institute), D-30167 Hannover, Germany}
% ...
\collaboration{(LIGO Scientific Collaboration and Virgo Collaboration)} 

\begin{abstract}
The Advanced LIGO and Advanced Virgo observatories recently discovered gravitational waves from a binary neutron star inspiral. A short gamma-ray burst (GRB) that followed the merger of this binary was also recorded by the Fermi Gamma-ray Burst Monitor (Fermi-GBM), and the Anticoincidence Shield for the Spectrometer for the International Gamma-Ray Astrophysics Laboratory (INTEGRAL), indicating particle acceleration by the source. The precise location of the event was determined by optical detections of emission following the merger. We searched for high-energy neutrinos from the merger in the GeV--EeV energy range using the {\sc Antares}, IceCube, and Pierre Auger Observatories. No neutrinos directionally coincident with the source were detected within $\pm500$\,s around the merger time. Additionally, no MeV neutrino burst signal was detected coincident with the merger. We further carried out an extended search in the direction of the source for high-energy neutrinos within the 14-day period following the merger, but found no evidence of emission. We used these results to probe dissipation mechanisms in relativistic outflows driven by the binary neutron star merger. The non-detection is consistent with model predictions of short GRBs observed at a large off-axis angle.
\end{abstract}


\section{Introduction}
\label{sec:introduction}

The observation of binary neutron star mergers with multiple cosmic messengers is a unique opportunity that enables the detailed study of the merger process, and provides insight into astrophysical particle acceleration and high-energy emission (e.g., \citealt{2012LRR....15....8F,2014ARA&A..52...43B,2013CQGra..30l3001B,GWGRB}). Binary neutron star mergers are prime sources of gravitational waves (GWs; e.g., \citealt{2010CQGra..27q3001A}), which provide information on the neutron star masses and spins (e.g., \citealt{2015PhRvD..91d2003V}). Kilonova/macronova observations of the mergers provide further information on the mass ejected by the disruption of the neutron stars (e.g., \citealt{2017LRR....20....3M,kilonova}).

Particle acceleration and high-energy emission by compact objects are currently not well understood (e.g., \citealt{2013APh....43..134M,2015PhR...561....1K}), and could be deciphered by combined information on the neutron star masses, ejecta mass, and gamma-ray burst (GRB) properties, as expected from multimessenger observations. In particular, the observation of high-energy neutrinos would reveal the hadronic content and dissipation mechanism in relativistic outflows \citep{PhysRevLett.78.2292}. A quasi-diffuse flux of high-energy neutrinos of cosmic origin has been identified by the IceCube observatory \add{\citep{2013PhRvL.111b1103A,2013Sci...342E...1I}}. The source population producing these neutrinos is currently not known.

On August 17, 2017, the Advanced LIGO \citep{2015CQGra..32g4001L} and Advanced Virgo \citep{2015CQGra..32b4001A} observatories recorded a GW signal, GW170817, from a binary neutron star inspiral \citep{BNS}.  Soon afterwards, Fermi-GBM and INTEGRAL detected a short GRB, GRB170817A, from a consistent location \citep{FermiDiscovery,INTEGRALDiscovery,GWGRB}. Subsequently, ultra-violet, optical, and infrared emission was observed from the merger, consistent with kilonova/macronova emission. Optical observations allowed the precise localization of the merger in the galaxy NGC\,4993, at equatorial coordinates
$\alpha{(\mathrm J2000.0)=13^{\mathrm h} 09^{\mathrm m}48\fs 085}$,
$\delta{(\mathrm J2000.0)}=-23^\circ 22\arcmin 53\farcs 343$ \citep{SwopeGCN,Swope,MMA}, and at a distance of $\sim$\,40\,Mpc. At later times, X-ray and radio emissions were also observed \citep{MMA}, consistent with the expected afterglow of a short GRB at high viewing angles (e.g., \citealt{GWGRB}).

High-energy neutrino observatories continuously monitor the whole sky or a large fraction of it, making them well suited to study emission from GW sources, even for unknown source locations or for emission prior to or after the GW detection \citep{2016PhRvD..93l2010A,2017PhRvD..96b2005A}. It is also possible to rapidly analyze the recorded data and inform other observatories in case of a coincident detection, significantly reducing the source localization uncertainty compared to that provided by GW information alone.

In this \emph{Letter} we present searches for high-energy neutrinos in coincidence with GW170817/GRB170817A by the three most sensitive high-energy neutrino observatories: (1) the {\sc Antares} neutrino telescope (hereafter {\sc Antares}; \citealt{2011NIMPA.656...11A}), a ten megaton-scale underwater Cherenkov neutrino detector located at a depth of 2500\,m in the Mediterranean Sea; (2) the IceCube Neutrino Observatory (hereafter IceCube; \citealt{Aartsen:2016_JINST}), a gigaton-scale neutrino detector installed 1500\,m deep in the ice at the geographic South Pole, Antarctica; and (3) the Pierre Auger Observatory (hereafter Auger; \citealt{Auger_observatory_NIMA15}), a cosmic-ray air-shower detector consisting of 1660 water-Cherenkov stations spread over an area of $\sim$\,3000\,km$^2$. All three detectors joined the low-latency multimessenger follow-up effort of LIGO-Virgo starting with LIGO's second observation run, O2.

Upon the identification of the GW signal GW170817, preliminary information on this event was rapidly shared with partner observatories \citep{MMA}. In response, IceCube \citep{IceCubeGCN1,IceCubeGCN2,IceCubeGCN3}, {\sc Antares} \citep{AntaresGCN1,AntaresGCN2}, and Auger \citep{AugerGCN} promptly searched for a neutrino counterpart, and shared their initial results with partner observatories. Subsequently, the three facilities carried out a more in-depth search for a neutrino counterpart using the precise localization of the source.

This \emph{Letter} is organized as follows. In Section \ref{sec:results}, we present the neutrino searches carried out by {\sc Antares}, IceCube, and Auger, as well as the results obtained. In Section \ref{sec:constraints}, we present constraints on processes in the merger that can lead to neutrino emission. We summarize our findings and conclude in Section \ref{sec:conclusion}.

\section{Searches and Results}
\label{sec:results}

\begin{figure*}
\begin{center}
\resizebox{0.8\textwidth}{!}{\includegraphics{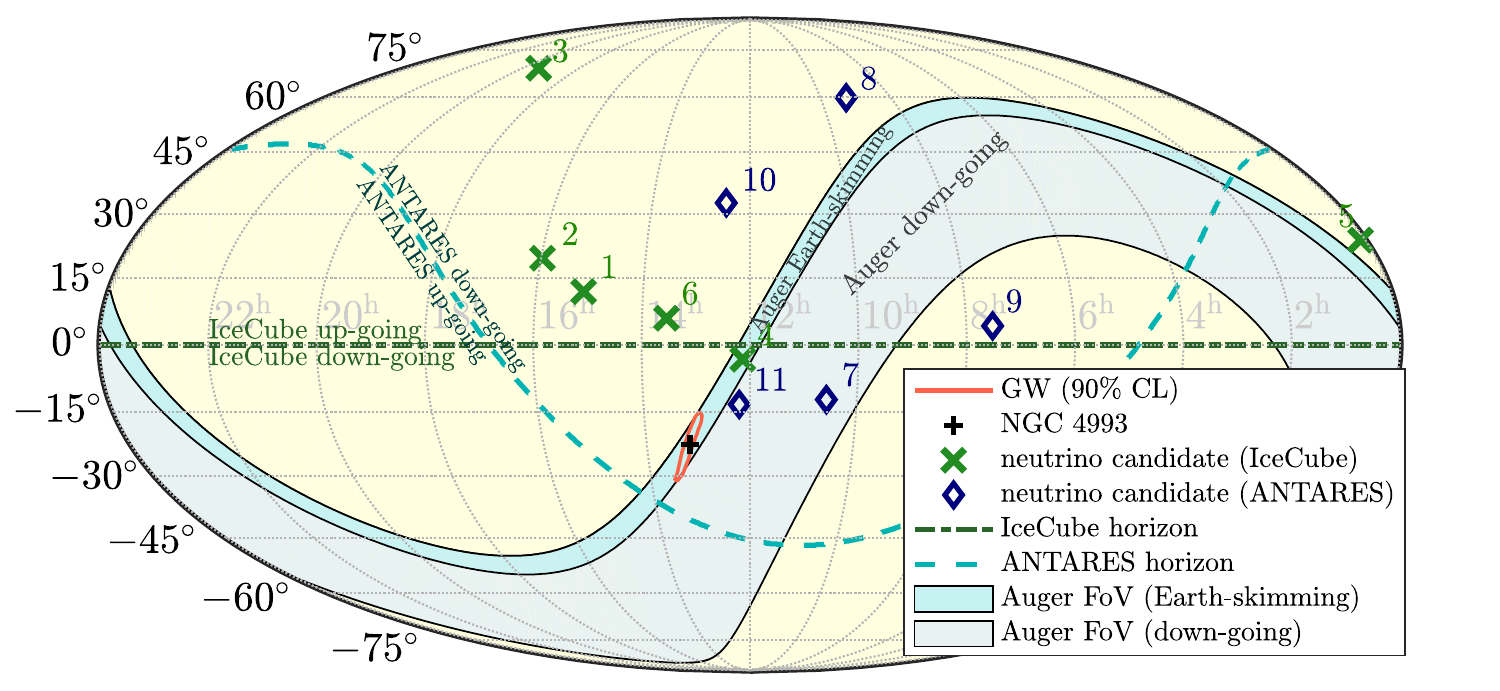}}
\end{center}
\caption{Localizations and sensitive sky areas at the time of the GW event in equatorial coordinates: GW 90\% credible-level localization (red contour; \citealt{BNS}), direction of NGC\,4993 (black plus symbol; \citealt{Swope}), directions of IceCube's and {\sc Antares}'s neutrino candidates within $500\,$s of the merger (green crosses and blue diamonds, respectively), {\sc Antares}'s horizon separating down-going (north of horizon) and up-going (south of horizon) neutrino directions (dashed blue line), and Auger's fields of view for Earth-skimming (darker blue) and down-going (lighter blue) directions. IceCube's up-going and down-going directions are on the northern and southern hemispheres, respectively. The zenith angle of the source at the detection time of the merger was $73.8^\circ$ for {\sc Antares}, $66.6^\circ$ for IceCube, and $91.9^\circ$ for Auger.}
\label{fig:skymap}
\end{figure*}

Neutrino observatories detect secondary charged particles produced in neutrino interaction with matter.
Surface detectors, such as Auger, use arrays of widely-spaced water Cherenkov detectors to observe the air-shower particles created by high-energy neutrinos.
In detectors such as {\sc Antares} and IceCube, three-dimensional arrays of optical modules deployed in water or ice detect the Cherenkov radiation from secondary charged particles that travel through the instrumented detector region.  For these detectors, the secondary particles can create two main event classes: track-like events from charged-current interactions of muon neutrinos and
from a minority of tau neutrino interactions; and shower-like events from all other interactions (neutral-current interactions and charged-current
interactions of electron and tau neutrinos). While energy deposition in track-like events can happen over distances of $\mathcal{O}$(km), shower-like events are confined to much smaller regions.

For all detectors, neutrino signals must be identified on top of a persistent background of charged particles produced by the interaction of cosmic ray particles
with the atmosphere above the detectors.  This discrimination is done by considering the observed direction and energy of the charged particles.  Surface detectors focus on high-energy ($\gtrsim$\,$10^{17}$eV)
showers created close to the detector by neutrinos from near-horizontal directions.  In-ice and in-water detectors can select well-reconstructed track events
from the up-going direction where the Earth is used as a natural shield
for the dominant background of penetrating muons from cosmic ray showers.
By requiring the neutrino interaction vertex to be contained inside the instrumented volume, or requiring its energy to be sufficiently high to be incompatible with the down-going muon background, even neutrino events originating above the horizon are identifiable.  Neutrinos originating from cosmic ray interactions in the atmosphere are also observed and constitute the primary background for up-going and vertex-contained event selections.

All three observatories, {\sc Antares}, IceCube, and Auger, performed searches for neutrino signals in coincidence with the binary neutron star merger
event GW170817, each using multiple event selections.  Two different time windows were used for the searches. First, we used a $\pm 500$\,s time window around the merger to search for neutrinos associated with prompt and extended gamma-ray emission \citep{2011APh....35....1B,2017ApJ...848L...4K}. Second, we searched for neutrinos over a longer 14-day time window following the GW detection, to cover predictions of longer-lived emission processes (e.g., \citealt{2013PhRvD..88d3010G,2017arXiv170704263F}).

%%%%%%%%%%%%%%%%%%%%%%%%%%%%%%%%%%%%%%%%%%%%
\subsection{ANTARES}
%%%%%%%%%%%%%%%%%%%%%%%%%%%%%%%%%%%%%%%%%%%%

%\color{blue}
The {\sc Antares} neutrino telescope has been continuously operating since 2008. Located deep (2500\,m) in the Mediterranean Sea, 40\,km from Toulon (France), it is a 10\,Mt-scale array of photosensors, detecting neutrinos with energies above $\mathcal{O}(100)$ GeV.

Based on the originally communicated locations of the GW signal and the GRB detection, high-energy neutrino candidates were initially searched for in the {\sc Antares} online data stream, relying on a fast algorithm which selects only up-going neutrino track candidates~\citep{AntaresOnline}. No up-going muon neutrino candidate events were found in a $\pm 500$\,s time window centered on the GW event time -- for an expected number of atmospheric background events of $\sim$\,$10^{-2}$ during the coincident time window. An extended online search during $\pm 1$\,h also resulted in no up-going neutrino coincidences.

As it subsequently became clear, the precise direction of origin of GW170817 in NGC\,4993 was above the {\sc Antares} horizon at the detection time of the binary merger (see Fig.\,\ref{fig:skymap}). Thus, a dedicated analysis looking for down-going muon neutrino candidates in the online {\sc Antares} data stream was also performed. No neutrino counterparts were found in this analysis. The results of these low-latency searches were shared with follow-up partners within a few hours for the up-going search and a few days for the down-going search \citep{AntaresGCN1,AntaresGCN2}.

Here, {\sc Antares} used an updated high-energy neutrino follow up of GW170817 that includes the shower channel. It was performed with the offline-reconstructed dataset, that incorporates dedicated calibration in terms of positioning, timing and efficiency~\citep{AntaresPosition,AntaresTime,AntaresEff}. The analysis has been optimized to increase the sensitivity of the detector and extended to the longer time window of 14 days.

The search for down-going neutrino counterparts to GW170817 was
made feasible as the large background affecting this dataset can be drastically suppressed by requiring a time and space coincidence with the GW signal. It was optimized, independently for tracks and showers, such that a directional coincidence with NGC\,4993 within the search time window of $\pm 500$\,s would have 3$\sigma$ significance. Muon neutrino candidates were selected by applying cuts on the estimated angular error and the track quality reconstruction parameter. \add{While {\sc Antares} is sensitive to neutrino events with energy as small as $\mathcal{O}(100\,\mbox{GeV})$, the energy range corresponding to the $5\%-95\%$ quantiles of the neutrino flux for a E$^{-2}$ signal spectrum is equal to [32\,TeV;\,22\,PeV]. For such a flux, the median angular uncertainty, defined as the median value of the distribution of angles between the reconstructed direction of the event and the true neutrino direction, is equal to $0.5^\circ$.}

Shower events were selected by applying a set of cuts primarily devoted to reducing the background rate~\citep{AntaresPS2017}. The energy range corresponding to the 5\%--95\% quantiles of the neutrino flux for a $E^{-2}$ signal spectrum is equal to [23 TeV; 16 PeV], while the median angular error is 6$^\circ$ with this set of relaxed cuts.

No events temporally coincident with GW170817 were found. Five background track events (likely atmospheric muons), not compatible with the source position, were detected (see Fig.\,\ref{fig:skymap}). We used this non-detection to constrain the neutrino fluence (see Fig. \ref{fig:limits}) which was computed as in \cite{2016PhRvD..93l2010A}.

The search over 14 days is restricted to up-going events, but includes all neutrino flavors (tracks and showers). We applied quality cuts optimized for point-source searches which give a median pointing accuracy of 0.4$^\circ$ and 3$^\circ$ respectively for track and shower events~\citep{AntaresPS2017}. No events spatially coincident with GRB170817A were found.

Compared to the upper limits obtained for the short time window of $\pm500$\,s, those limits are significantly less stringent above 1\,PeV, where the absorption of neutrinos by the Earth becomes important for up-going events. Below 10~TeV, the constraints computed for the 14-day time window are stricter due to the better acceptance in this energy range for up-going neutrino candidates compared to down-going events (see Fig.\,\ref{fig:limits}).

%%%%%%%%%%%%%%%%%%%%%%%%%%%%%%%%%%%%%%%%%%%%
\subsection{IceCube}
%%%%%%%%%%%%%%%%%%%%%%%%%%%%%%%%%%%%%%%%%%%%

IceCube is a cubic-kilometer size neutrino detector~\citep{Aartsen:2016_JINST}
installed in the ice at the geographic South Pole in Antarctica between depths of 1450\,m and 2450\,m.
Detector construction was completed in 2010, and the detector has operated with a $\sim$\,99\% duty cycle since.
IceCube searched for neutrino signals from GW170817 using two different event selection techniques.

The first search used an online selection of through-going muons, which is used in IceCube's online analyses \citep{1742-6596-718-6-062029,2016arXiv161206028I}
and follows an event selection similar to that of point source searches \citep{2014ApJ...796..109A}.
This event selection picks out primarily cosmic-ray-induced background events, with an expectation of 4.0 events in the
northern sky (predominantly generated by atmospheric neutrinos) and 2.7 events in the southern sky (predominantly muons generated by high energy cosmic rays interactions in the atmosphere
above the detector)
per 1000 seconds.  For source locations in the southern sky, the sensitivity of the down-going event selection for neutrinos below 1\,PeV weakens
rapidly with energy due to the rapidly increasing atmospheric muon background at lower energies.
Events found by this track selection in the $\pm 500$\,s time window are shown in Fig.\,\ref{fig:skymap}.  No
events were found to be spatially and temporally correlated with GW170817.

A second event selection, described in \cite{2017arXiv171001191I}, was employed offline. This uses the outermost optical sensors of the instrumented volume to veto incoming muon tracks from atmospheric background events. Above 60 TeV, this event selection has the same performance as the high-energy starting event selection~\citep{Aartsen:2014gkd}.  Below this energy, additional veto cuts similar to those described in \cite{Aartsen:2014muf} are applied, in order to maintain a low background level at energies down to a few TeV. Both track- and cascade-like events are retained. The event rate for this selection varies over the sky, but is overall much lower than for the online track selection described above. Between declinations $-13^{\circ}$ and $-33^\circ$, the mean number of events in a two-week period is 0.4 for tracks and 2.5 for cascades.  During the $\pm 500$\,s time-window, no events passed this event selection from anywhere in the sky.

A combined analysis of the IceCube through-going track selection and the starting-event selection allows upper limits to be placed on the neutrino fluence from GW170817 between the energies of 1 TeV and 1 EeV, shown in Fig.\,\ref{fig:limits}. In the central range from 10 TeV to 100 PeV, the upper limit for an $E^{-2}$ power-law spectral fluence is $F(E)=0.19\,(E/\mathrm{GeV})^{-2}\,\mathrm{GeV}^{-1}\,\mathrm{cm}^{-2}$.

Both the through-going track selection and the starting event selection were applied to data collected in the 14-day period following the time of GW170817.  Because of IceCube's location at the South Pole and 99.88\% on-time during the 14-day period, the exposure to the source location is continuous and unvaried. No spatially and temporally coincident events were seen in either selection during this follow-up period. The resulting upper limits are presented in Fig.\,\ref{fig:limits}.  At most energies these are unchanged from the short time-window. At the lowest energies, where most background events occur, the analysis effectively requires stricter criteria for a coincident event than were required in the short time window; the limits are correspondingly higher.  In the central range from 10 TeV to 100 PeV, the upper limit on an $E^{-2}$ power-law spectral fluence is $F(E)=0.23\times\left(E/\mathrm{GeV}\right)^{-2}\,\mathrm{GeV}^{-1}\,\mathrm{cm}^{-2}$.

The IceCube detector is also sensitive to outbursts of MeV neutrinos via a simultaneous increase in all photomultiplier signal rates. A neutrino burst signal from a galactic core-collapse supernova would be detected with high precision \citep{2011A&A...535A.109A}. The detector global dark rate is monitored continuously, the influence of cosmic ray muons is removed and low-level triggers are formed when deviations from the nominal rate exceed pre-defined levels. No alert was triggered during the $\pm 500$\,second time-window around the GW candidate. This is consistent with our expectations for cosmic events such as core-collapse supernovae or compact binary mergers that are significantly farther away than Galactic distances.

%%%%%%%%%%%%%%%%%%%%%%%%%%%%%%%%%%%%%%%%%%%%
\subsection{Pierre Auger Observatory}
%%%%%%%%%%%%%%%%%%%%%%%%%%%%%%%%%%%%%%%%%%%%

With the surface detector (SD) of the Pierre Auger
Observatory in Malarg\"ue, Argentina \citep{Auger_observatory_NIMA15},
air showers induced by ultra-high energy (UHE) neutrinos
can be identified for energies above $\sim$\,$10^{17}$ eV in the more numerous
background of UHE cosmic rays \citep{Auger_nu_PRD15}.
The SD consists of 1660 water-Cherenkov stations spread over an
area of $\sim$\,3000\,km$^2$ following a triangular arrangement of 1.5\,km grid spacing
\citep{Auger_observatory_NIMA15}.
The signals produced by the passage of shower particles through
the SD detectors are recorded as time traces in 25 ns intervals.

Cosmic rays interact shortly after entering
the atmosphere and induce extensive air showers.
For highly inclined directions their electromagnetic
component gets absorbed due to the large grammage
of atmosphere from the first interaction point to the ground.
As a consequence, the shower front at ground
level is dominated by muons that induce
sharp time traces in the water-Cherenkov stations. On the contrary,
showers induced by downward-going neutrinos at large zenith angles can start
their development deep in the atmosphere producing traces that spread over
longer times. These showers have a considerable fraction of electrons
and photons which undergo more interactions
than muons in the atmosphere, spreading more in time as they pass through the detector.
This is also the case for Earth-skimming showers, mainly
induced by tau neutrinos ($\nu_\tau$) that traverse horizontally below the
Earth's crust, and interact near the exit point inducing a tau lepton that escapes the Earth
and decays in flight in the atmosphere above the SD.

Dedicated and efficient selection criteria based on
the different time profiles of the signals detected in showers created by hadronic and neutrino primaries, enable the search for Earth-skimming as well as downward-going
neutrino-induced showers \citep{Auger_nu_PRD15}.
Deeply-starting downward-going showers initiated by neutrinos of any flavor
can be efficiently identified for zenith angles of $60^\circ < \theta < 90^\circ$ \citep{Auger_nu_PRD15}.
For the Earth-skimming channel typically only $\nu_\tau$-induced showers with zenith
angles $90^\circ < \theta < 95^\circ$ can trigger the SD.
This is the most sensitive channel to UHE neutrinos,
mainly due to the larger grammage and higher density of the target (the Earth)
where neutrinos are converted and where tau leptons can travel tens of kilometers \citep{Auger_nu_PRD15}.
The angular resolution of the Auger SD for inclined showers is better than $2.5^\circ$,
improving significantly as the number of triggered stations increases \citep{Auger_angular_resolution}.

%%%% RESULTS %%%%

Auger performed a search for UHE neutrinos with its SD in a time window of $\pm 500$\,s
centered at the merger time of GW170817 \citep{MMA},
as well as in a 14-day period after it \citep{2009PhRvD..79j3001M,2013PhRvD..88d3010G, 2017arXiv170704263F}.

The sensitivity to UHE neutrinos in Auger is limited to
large zenith angles, so that at each instant they
can be efficiently detected only from a specific fraction of
the sky \citep{Auger_point_sources_ApJL12, Auger_GW_PRD16}.
Remarkably, the position of the optical counterpart in NGC\,4993 \citep{SwopeGCN,Swope,MMA}
is visible
from Auger in the field of view of the Earth-skimming channel
during the whole $\pm 500$\,s window as shown in Fig.\,\ref{fig:skymap}.
In this time period the source of GW170817
transits from $\theta\sim$\,$93.3^\circ$ to $\theta\sim$\,$90.4^\circ$
as seen from the center of the array.
The performance of the Auger SD array (regularly monitored every
minute) is very stable in the $\pm 500$\,s window around GW170817,
with an average number of active stations
amounting to $\sim$\,$95.8 \pm 0.1\,\%$ of the 1660 stations
of the SD array.

No inclined showers passing the Earth-skimming selection (neutrino candidates)
were found in the time window $\pm 500$\,s around the trigger time of GW170817.
The estimated number of background events from cosmic rays in a 1000\,s period is
$\sim$\,$6.3 \times 10^{-7}$ for the cuts applied in the Earth-skimming analysis \citep{Auger_nu_PRD15}.

The absence of candidates in the $\pm 500$\,s window allows us to constrain
the fluence in UHE neutrinos from GW170817, assuming they are emitted
steadily in this interval and with an $E^{-2}$ spectrum \citep{Auger_GW_PRD16}.
Single-flavor differential limits to the spectral fluence are shown in Fig.~\ref{fig:limits},
in bins of one decade in energy. The sensitivity of the
observatory is largest in the energy bin around $10^{18}$\,eV.
The single-flavor upper limit to the spectral fluence is $F(E)=0.77\,(E/\mathrm{GeV})^{-2}\,\mathrm{GeV}^{-1}\,\mathrm{cm}^{-2}$
over the energy range from $10^{17}$\,eV to $2.5 \times 10^{19}$\,eV.

In the 14-day search period, as the Earth rotates, the position of NGC\,4993
transits through the field of view of the Earth-skimming and downward-going channels. As seen
from the Pierre Auger Observatory, the zenith angle of the optical counterpart oscillates daily between
$\theta\sim$\,$11^\circ$ and $\theta \sim$\,$121^\circ$. The source is visible in the Earth-skimming channel
for $\sim$\,$4\%$ of the day, and in the downward-going channel for $\sim$\,$10.5\,\%$ ($\sim$\,$11.1\,\%$)
in the zenith angle range $60^\circ < \theta < 75^\circ$ ($75^\circ < \theta < 90^\circ$).
No neutrino candidates were identified in the two-week search period.
Single-flavor differential limits to the spectral fluence are shown in Fig.~\ref{fig:limits}.
The corresponding upper limit to the spectral fluence is $F(E)=25\,
(E/\mathrm{GeV})^{-2}\,\mathrm{GeV}^{-1}\,\mathrm{cm}^{-2}$
over the same energy interval as for the $\pm 500$\,s time window, where the difference is due to the relatively long periods of time when
the source of GW170817 is not visible in the inclined directions.

\section{Discussion}
\label{sec:constraints}

\add{The nature of high-energy emission from the binary merger and its aftermath is not yet clear.} We compared the expected spectral fluence for different emission scenarios to our observational upper limits to probe the properties of the merger and its aftermath. \add{Here we briefly outline the relevant information from electromagnetic observations, and present our results for the different emission scenarios.}

The merger occurred at a distance of $\sim$\,$40$\,Mpc, which is the distance of its host galaxy NGC\,4993, identified through electromagnetic observations \citep{SwopeGCN,Swope,MMA}. The prompt gamma-ray emission from the source, GRB170817A, had an observed isotropic-equivalent energy of $E_{\rm iso}\approx 4\times10^{46}$\,erg, as recorded by Fermi-GBM \citep{GWGRB}. \add{This is orders of magnitude below typical observed short-GRB energies \citep{2014ARA&A..52...43B,GWGRB}.}

\add{Prompt gamma-ray emission in at least some short GRBs is followed by a weaker, extended emission that can last for hundreds of seconds \citep{2006ApJ...643..266N,2017ApJ...848L...4K}. Fermi-GBM did not detect a temporally extended emission following GRB170817A, placing a constraint of $\sim$\,$2\times10^{46}$\,erg\,s$^{-1}$ for a 10\,s long emission period over 1\,keV--10\,MeV \citep{GWGRB}, significantly below typical luminosities observed for extended emission.}

\add{The very faint gamma-ray emission, along with its observed, delayed afterglow, are consistent with a typical short GRB viewed off-axis \citep{2017arXiv171006421G,2017arXiv171005905I,2017arXiv171008514F,2017arXiv171005433T,Hallinaneaap9855,2017ApJ...848L..34M,2017ApJ...848L..23F,2041-8205-848-2-L20,2041-8205-848-2-L25,2017arXiv171005847K,2041-8205-848-2-L34}. A GRB is viewed off-axis if its viewing angle $\theta_{\rm obs}$, defined as the angle between the jet axis and the line of sight, is greater than the jet opening half-angle $\theta_j$ \citep{1538-4357-570-2-L61}. The viewing angle inferred from the data is $\theta_{\rm obs}\gtrsim20^\circ$, while typical opening half-angles for short GRBs are within $\theta_{\rm j}\approx3^\circ-10^\circ$ \citep{2014ARA&A..52...43B}.}

GW data combined with the measured redshift of the host galaxy further provide constraints on $\theta_{\rm obs}$. Here we assume that the jet axis is aligned with the binary's total angular momentum vector. Adopting the Hubble constant from cosmic microwave background measurements by the Planck satellite \citep{2016A&A...594A..13P}, these data are consistent with  $\theta_{\rm obs}=0$, but also allow for a misalignment of $\theta_{\rm obs}\leq\PlanckHnaughtViewingNinetyULTidalNoSpins^\circ$ at $90\%$ credible level. Adopting the Hubble constant from Type Ia supernova measurements \citep{2016ApJ...826...56R} gives a similar result with maximum misalignment of $\theta_{\rm obs}\leq\RiessHnaughtViewingNinetyULTidalNoSpins^\circ$ at $90\%$ credible level \citep{H0}.

\begin{figure}
\begin{center}
\resizebox{0.46\textwidth}{!}{\includegraphics{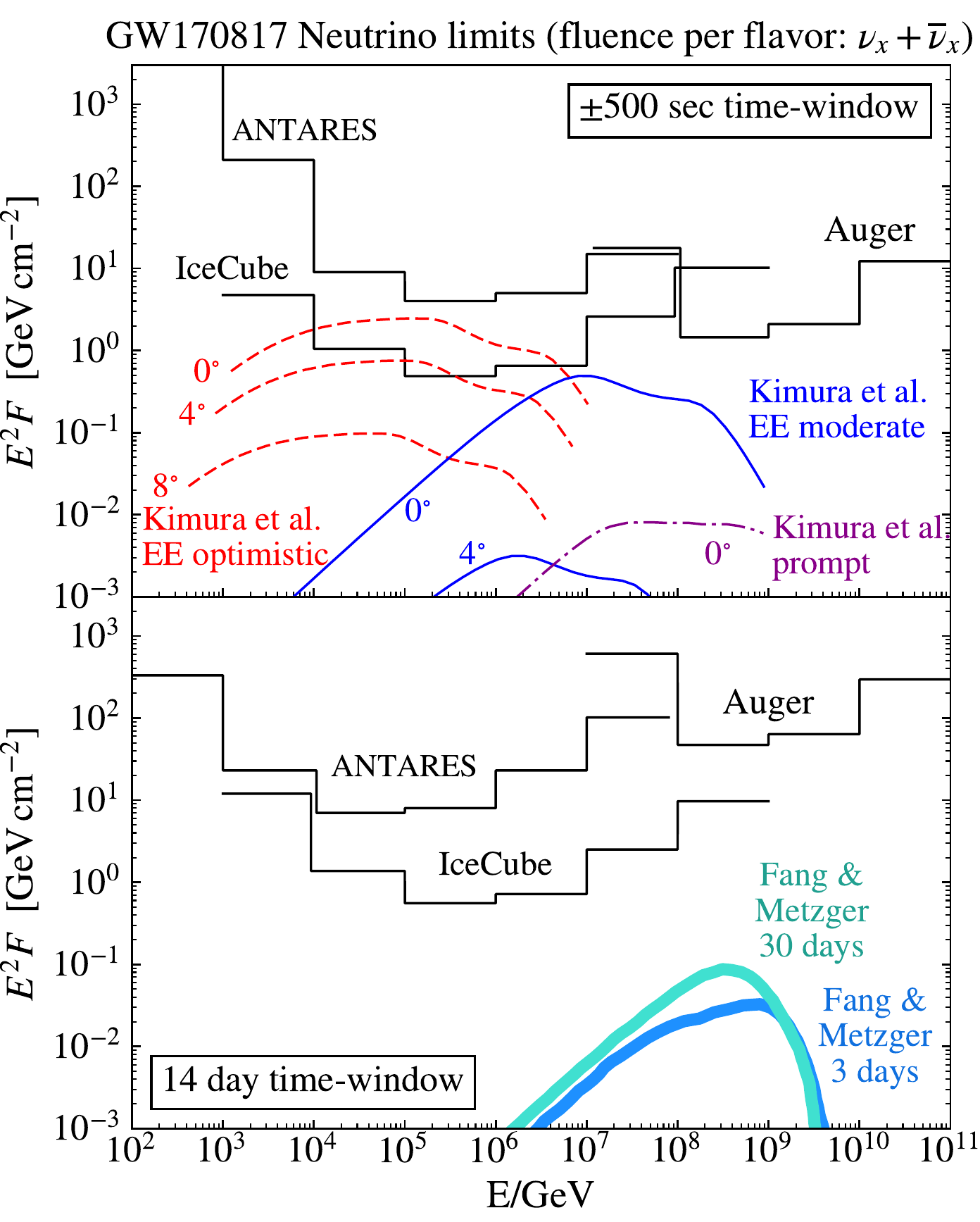}}
\end{center}
\caption{Upper limits (at $90\,\%$ confidence level) on the neutrino spectral fluence from GW170817 during a $\pm 500$\,s window centered on the GW trigger time (top panel), and a 14-day window following the GW trigger (bottom panel). For each experiment, limits are calculated separately for each energy decade, assuming a spectral fluence
$F(E)=F_\mathrm{up}\times\left[E/\mathrm{GeV}\right]^{-2}$ in that decade only. Also shown are predictions by neutrino emission models. In the upper plot, models from \cite{2017ApJ...848L...4K} for both extended emission (EE) and prompt emission are scaled to a distance of 40~Mpc, and shown for the case of on-axis viewing angle \add{($\theta_{\rm obs} \lesssim \theta_j$)} and selected off-axis angles to indicate the dependence on this parameter. \add{The shown off-axis angles are measured in excess of the jet opening half angle $\theta_j$.} GW data and the redshift of the host-galaxy constrain the viewing angle to $\theta_{\rm obs}\in[0^\circ,\RiessHnaughtViewingNinetyULTidalNoSpins^\circ]$ (see Section \ref{sec:constraints}). In the lower plot, models from \cite{2017arXiv170704263F} are scaled to a distance of 40\,Mpc.  All fluences are shown as the per flavor sum of neutrino and anti-neutrino fluence, assuming equal fluence in all flavors, as expected for standard neutrino oscillation parameters.
}
\label{fig:limits}
\end{figure}

\add{Considering this off-axis scenario, we examined the expected high-energy neutrino emission from a typical GRB observed at different viewing angles. The most promising neutrino production mechanism from GRBs is related to the extended gamma emission, due to its relatively low Lorentz factor resulting in high meson production efficiency \citep{2017ApJ...848L...4K}. In Fig.\,\ref{fig:limits} we compared our observational constraints with the expected neutrino fluence from the GRB's extended emission. For the on-axis (i.e. $\theta_{\rm obs} \lesssim \theta_j$) spectral fluence $F_{\rm on}$, we assumed the results of \cite{2017ApJ...848L...4K}, rescaled to 40\,Mpc. We approximated the observed off-axis spectral fluence, $F_{\rm off}(E)$ for these models using $F_{\rm off}(E) = \eta F_{\rm on}(E/\eta)$, where the scaling factor $\eta = \delta(\theta_{\rm obs})/\delta(0)$ accounts for different Doppler factors $\delta(\theta_{\rm obs}) = [\Gamma(1-\beta\cos(\theta_{\rm obs} - \theta_j))]^{-1}$ \citep{1538-4357-570-2-L61}.}

\add{For comparison, we also examined the expected neutrino flux associated with the prompt GRB emission (e.g., ]\citealt{2016PhRvD..93l3011M,2017ApJ...848L...4K}). This emission phase is less favorable for neutrino production than the extended emission. We see in Fig.\,\ref{fig:limits} that prompt emission from a single merger event is unlikely to produce a detected neutrino for the considered observatories, even if viewed on-axis.}

\add{Another proposed explanation for the faintness of gamma-ray emission is the interaction of the GRB jet with ejecta material from the merger \citep{Kasliwaleaap9455,2017arXiv171005896G,2017arXiv171005822P}. Energy deposition by the jet into the neutron star ejecta can form a cocoon that expands outwards at mildly relativistic speeds over a wide opening angle. Faint gamma-ray emission is then expected during the breakout of this cocoon from the outer tail of the ejecta \citep{2017arXiv171005896G}.}

\add{High-energy neutrino production in this scenario may significantly exceed the observed gamma-ray emission as neutrinos can escape through the ejecta even before it becomes transparent to gamma-rays. This scenario resembles that of a jet burrowing through the stellar envelope in a core-collapse event \citep{PhysRevLett.87.171102,2003PhRvD..68h3001R,2012PhRvD..86h3007B,2013PhRvL.111l1102M}. Nevertheless, if the observed gamma-rays come from the outbreak of a wide cocoon, it is less likely that the relativistic jet, which is more narrowly beamed than the cocoon outbreak, also pointed towards Earth.}

We further considered an additional neutrino-production mechanism related to ejecta material from the merger. If a rapidly rotating neutron star forms in the merger and does not immediately collapse into a black hole, it can power a relativistic wind with its rotational energy, which may be responsible for the sometimes observed extended emission \citep{2008MNRAS.385.1455M}. Optically thick ejecta from the merger can attenuate the gamma-ray flux, while allowing the escape of high-energy neutrinos. Additionally, it may trap some of the wind energy until it expands and becomes transparent. This process can convert some of the wind energy to high-energy particles, producing a \emph{long-term} neutrino radiation that can last for days \citep{2009PhRvD..79j3001M,2013PhRvD..88d3010G,2017arXiv170704263F}. The properties of ejecta material around the merger can be characterized from its kilonova/macronova emission.

Considering the possibility that the relative weakness of gamma-ray emission from GRB170817A may be partly due to attenuation by the ejecta, we compared our neutrino constraints to neutrino emission expected for typical GRB parameters. For the prompt and extended emissions, we used the results of \cite{2017ApJ...848L...4K} and compared these to our constraints for the relevant $\pm500$\,s time window. For extended emission we considered source parameters corresponding to both optimistic and moderate scenarios in Table\,1 of \cite{2017ApJ...848L...4K}. For emission on even longer timescales, we compared our constraints for the 14-day time window with the relevant results of \cite{2017arXiv170704263F}, namely emission from approximately 0.3 to 3 days and from 3 to 30 days following the merger. Predictions based on fiducial emission models and neutrino constraints are shown in Fig.\,\ref{fig:limits}. We found that our limits would constrain the optimistic extended-emission scenario for a typical GRB at $\sim$\,$40$\,Mpc, viewed at zero viewing angle.

\section{Conclusion}
\label{sec:conclusion}

%The discovery of GW170817 and its electromagnetic counterparts represent the beginning of multimessenger astronomy.

We searched for high-energy neutrinos from the first binary neutron star merger detected through GWs, GW170817, in the energy band of [$\sim$\,$10^{11}$\,eV, $\sim$\,$10^{20}$\,eV] using the {\sc Antares}, IceCube, and Pierre Auger Observatories, as well as for MeV neutrinos with IceCube. This marks an unprecedented joint effort of experiments sensitive to high-energy neutrinos. We have observed no significant neutrino counterpart within a $\pm 500$\,s window, nor in the subsequent 14\,days. The three detectors complement each other in the energy bands in which they are most sensitive (see Fig.\,\ref{fig:limits}).

This non-detection is consistent with our expectations from a typical GRB observed off-axis, or with a low-luminosity GRB.  Optimistic scenarios for on-axis gamma-attenuated emission are constrained by the present non-detection. %Nevertheless, the possible large viewing angle of the GRB, consistent with GW constraints and the observed low gamma-ray brightness expected for typical GRB emission viewed off-axis, make the off-axis scenario a more plausible explanation than attenuation.
%\add{If the GRB is produced by cocoon shock breakout, it could mean stronger neutrino emission.}

%The IceCube and {\sc Antares} constraints apply in the energy range [$\sim 10^{11}$\,eV, $\sim 10^{17}$\,eV], while Auger's apply in the complementary energy range [$\sim 10^{17}$\,eV, $\sim 2.5 \times 10^{19}\,$eV], meaning that 90\% of the signal detectable by Auger is expected in this energy range assuming a power-law $dN_\nu/dE_\nu \sim E_\nu^{-2}$ neutrino flux. As the source was below the horizon at the time of detection of the event, the Auger limit for the 1000\,s period applies only to tau-flavored neutrinos as the probability of detecting ES electron or muon neutrinos is negligible.

While the location of this source was nearly ideal for Auger, it was well above the horizon for IceCube and {\sc Antares} for prompt observations. This limited the sensitivity of the latter two detectors, particularly below $\sim$\,$100$\,TeV. For source locations near, or below the horizon, a factor of $\sim$\,$10$ increase in fluence sensitivity to prompt emission from an $E^{-2}$ neutrino spectrum is expected.  %We look forward to future binary neutron star observations from the LIGO-Virgo detector in the future.

With the discovery of a nearby binary neutron star merger, the ongoing enhancement of detector sensitivity \citep{2016LRR....19....1A} and the growing network of GW detectors \citep{PhysRevD.88.043007,India}, we can expect that several binary neutron star mergers will be observed in the near future. Not only will this allow stacking analyses of neutrino emission, but it will also bring about sources with favorable orientation and direction.

% ANTARES NOT HERE!!
The {\sc Antares}, IceCube, and Pierre Auger Collaborations are planning to continue the rapid search for neutrino candidates from identified GW sources. A coincident neutrino, with a typical position uncertainty of $\sim$\,$1$\,deg$^2$ could significantly improve the fast localization of joint events compared to the GW-only case. In addition, the first joint GW and high-energy neutrino discovery might thereby be known to the wider astronomy community within minutes after the event, opening a rich field of multimessenger astronomy with particle, electromagnetic, and gravitational waves combined.

\begin{acknowledgments}
\section*{Acknowledgments}
% ANTARES %%%%%%%%%%%%%%%%%%%%%%%%%%%%%%%%%%%%%%%%%
{\bf (ANTARES)} The {\sc{Antares}} authors acknowledge the financial support of the funding agencies:
% France:
Centre National de la Recherche Scientifique (CNRS), Commissariat \`a
l'\'ener\-gie atomique et aux \'energies alternatives (CEA),
Commission Europ\'eenne (FEDER fund and Marie Curie Program),
Institut Universitaire de France (IUF), IdEx program and UnivEarthS
Labex program at Sorbonne Paris Cit\'e (ANR-10-LABX-0023 and
ANR-11-IDEX-0005-02), Labex OCEVU (ANR-11-LABX-0060) and the
A*MIDEX project (ANR-11-IDEX-0001-02),
R\'egion \^Ile-de-France (DIM-ACAV), R\'egion
Alsace (contrat CPER), R\'egion Provence-Alpes-C\^ote d'Azur,
D\'e\-par\-tement du Var and Ville de La
Seyne-sur-Mer, France;
% Germany:
Bundesministerium f\"ur Bildung und Forschung
(BMBF), Germany;
% Italy
Istituto Nazionale di Fisica Nucleare (INFN), Italy;
% Netherlands
Nederlandse organisatie voor Wetenschappelijk Onderzoek (NWO), the Netherlands;
% Russia
Council of the President of the Russian Federation for young
scientists and leading scientific schools supporting grants, Russia;
% Romania
National Authority for Scientific Research (ANCS), Romania;
% Spain
Mi\-nis\-te\-rio de Econom\'{\i}a y Competitividad (MINECO):
Plan Estatal de Investigaci\'{o}n (refs. FPA2015-65150-C3-1-P, -2-P and -3-P, (MINECO/FEDER)), Severo Ochoa Centre of Excellence and MultiDark Consolider (MINECO), and Prometeo and Grisol\'{i}a programs (Generalitat
Valenciana), Spain;
% Marocco
Ministry of Higher Education, Scientific Research and Professional Training, Morocco.
% A.O.B.:
We also acknowledge the technical support of Ifremer, AIM and Foselev Marine
for the sea operation and the CC-IN2P3 for the computing facilities.

% ICECUBE %%%%%%%%%%%%%%%%%%%%%%%%%%%%%%%%%%%%%%%%%
{\bf (IceCube)} The IceCube collaboration acknowledges the support from the following agencies:
U.S. National Science Foundation-Office of Polar Programs,
U.S. National Science Foundation-Physics Division,
University of Wisconsin Alumni Research Foundation,
the Grid Laboratory Of Wisconsin (GLOW) grid infrastructure at the University of Wisconsin - Madison, the Open Science Grid (OSG) grid infrastructure;
U.S. Department of Energy, and National Energy Research Scientific Computing Center,
the Louisiana Optical Network Initiative (LONI) grid computing resources;
Natural Sciences and Engineering Research Council of Canada,
WestGrid and Compute/Calcul Canada;
Swedish Research Council,
Swedish Polar Research Secretariat,
Swedish National Infrastructure for Computing (SNIC),
and Knut and Alice Wallenberg Foundation, Sweden;
German Ministry for Education and Research (BMBF),
Deutsche Forschungsgemeinschaft (DFG),
Helmholtz Alliance for Astroparticle Physics (HAP),
Initiative and Networking Fund of the Helmholtz Association,
Germany;
Fund for Scientific Research (FNRS-FWO),
FWO Odysseus programme,
Flanders Institute to encourage scientific and technological research in industry (IWT),
Belgian Federal Science Policy Office (Belspo);
Marsden Fund, New Zealand;
Australian Research Council;
Japan Society for Promotion of Science (JSPS);
the Swiss National Science Foundation (SNSF), Switzerland;
National Research Foundation of Korea (NRF);
Villum Fonden, Danish National Research Foundation (DNRF), Denmark.

% AUGER %%%%%%%%%%%%%%%%%%%%%%%%%%%%%%%%%%%%%%%%%%%
\begin{sloppypar}
{\bf (Auger)} The successful installation, commissioning, and operation of the Pierre
Auger Observatory would not have been possible without the strong
commitment and effort from the technical and administrative staff in
Malarg\"ue. We are very grateful to the following agencies and
organizations for financial support:
\end{sloppypar}

\begin{sloppypar}
Argentina -- Comisi\'on Nacional de Energ\'\i{}a At\'omica; Agencia Nacional de
Promoci\'on Cient\'\i{}fica y Tecnol\'ogica (ANPCyT); Consejo Nacional de
Investigaciones Cient\'\i{}ficas y T\'ecnicas (CONICET); Gobierno de la
Provincia de Mendoza; Municipalidad de Malarg\"ue; NDM Holdings and Valle
Las Le\~nas; in gratitude for their continuing cooperation over land
access; Australia -- the Australian Research Council; Brazil -- Conselho
Nacional de Desenvolvimento Cient\'\i{}fico e Tecnol\'ogico (CNPq);
Financiadora de Estudos e Projetos (FINEP); Funda\c{c}\~ao de Amparo \`a
Pesquisa do Estado de Rio de Janeiro (FAPERJ); S\~ao Paulo Research
Foundation (FAPESP) Grants No.~2010/07359-6 and No.~1999/05404-3;
Minist\'erio da Ci\^encia, Tecnologia, Inova\c{c}\~oes e Comunica\c{c}\~oes (MCTIC);
Czech Republic -- Grant No.~MSMT CR LG15014, LO1305, LM2015038 and
CZ.02.1.01/0.0/0.0/16\_013/0001402; France -- Centre de Calcul
IN2P3/CNRS; Centre National de la Recherche Scientifique (CNRS); Conseil
R\'egional Ile-de-France; D\'epartement Physique Nucl\'eaire et Corpusculaire
(PNC-IN2P3/CNRS); D\'epartement Sciences de l'Univers (SDU-INSU/CNRS);
Institut Lagrange de Paris (ILP) Grant No.~LABEX ANR-10-LABX-63 within
the Investissements d'Avenir Programme Grant No.~ANR-11-IDEX-0004-02;
Germany -- Bundesministerium f\"ur Bildung und Forschung (BMBF); Deutsche
Forschungsgemeinschaft (DFG); Finanzministerium Baden-W\"urttemberg;
Helmholtz Alliance for Astroparticle Physics (HAP);
Helmholtz-Gemeinschaft Deutscher Forschungszentren (HGF); Ministerium
f\"ur Innovation, Wissenschaft und Forschung des Landes
Nordrhein-Westfalen; Ministerium f\"ur Wissenschaft, Forschung und Kunst
des Landes Baden-W\"urttemberg; Italy -- Istituto Nazionale di Fisica
Nucleare (INFN); Istituto Nazionale di Astrofisica (INAF); Ministero
dell'Istruzione, dell'Universit\'a e della Ricerca (MIUR); CETEMPS Center
of Excellence; Ministero degli Affari Esteri (MAE); Mexico -- Consejo
Nacional de Ciencia y Tecnolog\'\i{}a (CONACYT) No.~167733; Universidad
Nacional Aut\'onoma de M\'exico (UNAM); PAPIIT DGAPA-UNAM; The Netherlands
-- Ministerie van Onderwijs, Cultuur en Wetenschap; Nederlandse
Organisatie voor Wetenschappelijk Onderzoek (NWO); Stichting voor
Fundamenteel Onderzoek der Materie (FOM); Poland -- National Centre for
Research and Development, Grants No.~ERA-NET-ASPERA/01/11 and
No.~ERA-NET-ASPERA/02/11; National Science Centre, Grants
No.~2013/08/M/ST9/00322, No.~2013/08/M/ST9/00728 and No.~HARMONIA
5--2013/10/M/ST9/00062, UMO-2016/22/M/ST9/00198; Portugal -- Portuguese
national funds and FEDER funds within Programa Operacional Factores de
Competitividade through Funda\c{c}\~ao para a Ci\^encia e a Tecnologia
(COMPETE); Romania -- Romanian Authority for Scientific Research ANCS;
CNDI-UEFISCDI partnership projects Grants No.~20/2012 and No.~194/2012
and PN 16 42 01 02; Slovenia -- Slovenian Research Agency; Spain --
Comunidad de Madrid; Fondo Europeo de Desarrollo Regional (FEDER) funds;
Ministerio de Econom\'\i{}a y Competitividad; Xunta de Galicia; European
Community 7th Framework Program Grant No.~FP7-PEOPLE-2012-IEF-328826;
USA -- Department of Energy, Contracts No.~DE-AC02-07CH11359,
No.~DE-FR02-04ER41300, No.~DE-FG02-99ER41107 and No.~DE-SC0011689;
National Science Foundation, Grant No.~0450696; The Grainger Foundation;
Marie Curie-IRSES/EPLANET; European Particle Physics Latin American
Network; European Union 7th Framework Program, Grant
No.~PIRSES-2009-GA-246806; European Union's Horizon 2020 research and
innovation programme (Grant No.~646623); and UNESCO.
\end{sloppypar}

% LIGO-Virgo %%%%%%%%%%%%%%%%%%%%%%%%%%%%%%%%%%%%%%
{\bf (LIGO and Virgo)} The authors gratefully acknowledge the support of the United States
National Science Foundation (NSF) for the construction and operation of the
LIGO Laboratory and Advanced LIGO as well as the Science and Technology Facilities Council (STFC) of the
United Kingdom, the Max-Planck-Society (MPS), and the State of
Niedersachsen/Germany for support of the construction of Advanced LIGO
and construction and operation of the GEO600 detector.
Additional support for Advanced LIGO was provided by the Australian Research Council.
The authors gratefully acknowledge the Italian Istituto Nazionale di Fisica Nucleare (INFN),
the French Centre National de la Recherche Scientifique (CNRS) and
the Foundation for Fundamental Research on Matter supported by the Netherlands Organisation for Scientific Research,
for the construction and operation of the Virgo detector
and the creation and support  of the EGO consortium.
The authors also gratefully acknowledge research support from these agencies as well as by
the Council of Scientific and Industrial Research of India,
the Department of Science and Technology, India,
the Science \& Engineering Research Board (SERB), India,
the Ministry of Human Resource Development, India,
the Spanish  Agencia Estatal de Investigaci\'on,
the  Vicepresid\`encia i Conselleria d'Innovaci\'o, Recerca i Turisme and the Conselleria d'Educaci\'o i Universitat del Govern de les Illes Balears,
the Conselleria d'Educaci\'o, Investigaci\'o, Cultura i Esport de la Generalitat Valenciana,
the National Science Centre of Poland,
the Swiss National Science Foundation (SNSF),
the Russian Foundation for Basic Research,
the Russian Science Foundation,
the European Commission,
the European Regional Development Funds (ERDF),
the Royal Society,
the Scottish Funding Council,
the Scottish Universities Physics Alliance,
the Hungarian Scientific Research Fund (OTKA),
the Lyon Institute of Origins (LIO),
the National Research, Development and Innovation Office Hungary (NKFI),
the National Research Foundation of Korea,
Industry Canada and the Province of Ontario through the Ministry of Economic Development and Innovation,
the Natural Science and Engineering Research Council Canada,
the Canadian Institute for Advanced Research,
the Brazilian Ministry of Science, Technology, Innovations, and Communications,
the International Center for Theoretical Physics South American Institute for Fundamental Research (ICTP-SAIFR),
the Research Grants Council of Hong Kong,
the National Natural Science Foundation of China (NSFC),
the Leverhulme Trust,
the Research Corporation,
the Ministry of Science and Technology (MOST), Taiwan
and
the Kavli Foundation.
The authors gratefully acknowledge the support of the NSF, STFC, MPS, INFN, CNRS and the
State of Niedersachsen/Germany for provision of computational resources.

%%%%%%%%%%%%%%%%%%%%%%%%%%%%%%%%%%%%%%%%%%%%%%%%%%%
\end{acknowledgments}

%#################################################################
%\bibliography{Refs}

  % Restore the author, affiliation and maketitle commands.
  %

%%%%%%%%%%%%%%%%%%%%%%%%%%%%%%%%%%%%%%%%%%%%%%%%%%%%%%%%%%%%%%%%%%%%%%%%%%%%%%
%  % Restore the author, affiliation and maketitle commands.
%  %
%  \let\author\myauthor
%  \let\affiliation\myaffiliation
%  \let\maketitle\mymaketitle
%  \title{Authors}
%  \pacs{}
%   \include{author_list_GWHEN_new_format1}
%   \include{IceCube_authorlist}
%   \include{Auger_aastex_authorlist.tex}
%   \include{LVC_authorlist}
%  \newpage
%  \maketitle
%%%%%%%%%%%%%%%%%%%%%%%%%%%%%%%%%%%%%%%%%%%%%%%%%%%%%%%%%%%%%%%%%%%%%%%%%%%%%%

\end{document}